\documentclass[12pt]{article}
\usepackage[latin1]{inputenc} 
\usepackage[dvips]{graphicx}
\usepackage[usenames,dvipsnames]{color}
\title{Nucleotide Frequencies in Human Genome and Fibonacci Numbers}
\author{Michel E. Beleza Yamagishi\thanks{Corresponding author: michel@cbi.cnptia.embrapa.br} \\ Embrapa Inform\'atica /UNISAL    \and Alex Itiro Shimabukuro \\PUC Campinas}

\begin{document}

\maketitle

\begin{quotation}
\begin{flushright}
	$ \Pi\rho\alpha\gamma\mu\alpha$  $\epsilon\iota\nu\alpha\iota$  $\alpha\rho\iota\theta\mu o \iota$
\end{flushright}
\end{quotation}
 
\begin{abstract}
{\em This work presents a mathematical model that establishes
an interesting connection between nucleotide 
frequencies in  human single-stranded DNA and the famous Fibonacci's 
numbers. The model relies on two assumptions. First, Chargaff's second 
parity rule should be valid, and, second, the nucleotide frequencies
should approach limit values when the number of bases is sufficiently
large. Under these two hypotheses, it is possible to predict
the human nucleotide frequencies with accuracy.
It is noteworthy, that the predicted values are solutions
of an optimization problem, which is commonplace 
in many nature's phenomena. }
\end{abstract}

\section{Introduction}

The amount of available genome data is increasing
very fast due the completion of a host of genome
sequencing projects. The careful analysis of 
all these data is only beginning. The genome
sequence by itself is meaningless, it is necessary 
to identify genes, proceed the annotation, and,
if possible, get some understanding of the very 
process responsible by the sequence formation.

Less than $ 25 \% $ of the fly genome is in coding regions,
and the number falls to less than $ 3 \% $ in humans \cite{Do}.
It seems that the most part of eukaryotes genomes is
``garbage'' DNA. Nevertheless, recently, some evidences
show that it is not the case. Mutations in noncoding
regions were associated with cancer \cite{schwartz}. Consequently,
the interest in noncoding regions has increased, and
the role that those regions have in the whole genome
demands a better comprehension.

The initial step in any genome analysis is to
perform some simple statistical measures like
frequencies and averages. These
kind of research have been done even before
the discovery of DNA structure, and allowed some
striking scientific advances. For instance, in 1951, 
Chargaff \cite{chargaff1} observed that,
in any piece of double-stranded DNA, the frequencies 
of adenine and thymine are equal, and so are the 
frequencies of cytosine and guanine. In mathematical notation $
P_A=P_T $ and $ P_C=P_G $, where $ P_A, P_C, P_G $ and $ P_T $
denote the nucleotide frequencies of adenine, cytosine, guanine and
thymine, respectively. This observation is known as
{\em Chargaff's first parity rule}. Watson and Crick, in 1953,
were acquainted with Chargaff's first parity rule, and
used it to support their DNA double-helix
model \cite{doublehelix}. Furthermore, Chargaff also observed
that the parity rule approximately holds
in a single-stranded DNA, nonetheless the equality is not strict,
but $ P_A \cong P_T $ and $ P_C \cong P_G $. 
This is known as {\em Chargaff's second parity 
rule}. Possibly, the best explanation to this rule
can be found in \cite{forsdyke}. Chargaff's second rule 
has been extensively tested \cite{mitchell} and proved to
hold in the majority of the genome sequences.
   
\begin{figure}[ht]
	\begin{center}
	\scalebox{0.8} % h_lenght
	{
		\includegraphics{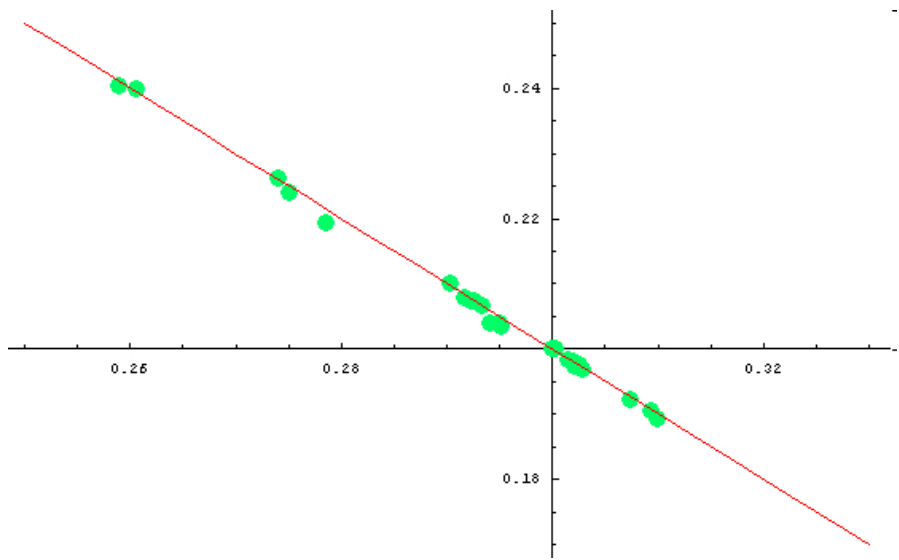}
  }
	\end{center}
	\caption{In red the line $ P_A + P_C = \frac{1}{2} $, and in green 
	the observed points $ (P_A,P_C) $ for each human chromosome}
	\label{fig:reta}
\end{figure}

A particular interesting case is found in human genome. We have
tested the Chargaff's second parity rule for each one
of the 24 human chromosomes $ (22+X+Y) $, and it
is definitely valid. Moreover, notice that $ P_A+P_T+P_C+P_G=1 $ 
by definition (the sum of all frequencies must be equal to 1), 
and  assuming {\em Chargaff's second parity  rule},
we get that $ P_A + P_C \cong \frac{1}{2} $ or, equivalently,
$P_T+P_G \cong \frac{1}{2} $, or any possible combination.
If we plot the points $ (P_A,P_C) $ for each
human chromosome, we get another interesting fact: they are not evenly spread
over the line $ P_A + P_C = \frac{1}{2}$, but seem to 
be aggregated around some very precise values. In
Figure \ref{fig:reta}, in red, the line $ P_A + P_C = \frac{1}{2}$, 
and the green dots are the points $ (P_A,P_C) $ for each  human chromosome.

Although this observation is not expected, it is not completely
unusual. Many phenomena in nature show the same pattern, and some 
of them can be mathematically modeled. Usually, those mathematical
models that describe nature's phenomena involve optimization problems.
It seems that nature is always trying to optimize itself in different
contexts. Therefore, the following question emerges naturally:
Is it possible to build a mathematical model that predicts or explain the 
observed frequencies? 

Assuming that (i) the human nucleotide frequencies really tend to limit values
when the number os bases is sufficiently large, and (ii) Chargaff's second
parity rule is valid, we derived a mathematical model
that predicts the observed frequency values with accuracy.

\section{Mathematical Model}

 In order to understand our model, it is necessary to 
 introduce the Fibonacci numbers \cite{fibo}. 
 
 \subsection{Fibonacci Numbers}
 
 In mathematics, one of the most famous integer sequence
 is without doubt the sequence $ \{ 1,1,2,3,5,8,13,...\}$.
 
 This sequence, called Fibonacci sequence, is obtained through the recurrence formula
 
 \begin{equation}
 \label{eq1}
 F(n+2)=F(n+1)+F(n), 
 \end{equation}
 together with the initial conditions  $ F(1)=1 $ and $ F(2)=1 $. 
 
 The Fibonacci sequence was first described, in the Occident, by {\em Leonardo of Pisa},
 also known as Fibonacci, in his book {\em Liber Abaci}. The Fibonacci
 sequence appears in nature in different contexts: sea shell shapes,
 flower petals and seeds, etc. 
 
 It is related to the {\em Golden Ratio}, $ \phi $, by the limit 
 
 \begin{equation}
 \label{eq1.1}
 \phi=\lim_{n\rightarrow\infty}\frac{F(n+1)}{F(n)}.
 \end{equation}
 
 The Golden Ratio is associated to Beauty and Perfection, and for this
 reason it is conventional to find $ \phi$ present in art (Leonardo da Vinci), 
 architecture (Parthenon in Athens, for example) and music 
 (notably in Bart\'ok and Debussy). There is a plenty of written
 works about the Golden Ratio and the Fibonacci numbers.
 
\subsection{Assumptions and Model}

 The main assumption of our model is that Chargaff's second parity
 rule is valid in all human chromosomes. There are many different forms
 to state it mathematically. We've decided to do it in the following way: {\em the 
 division of the frequency of one nucleotide by the sum of the frequencies 
 of the remaining nucleotides is in the proportion of  three 
 Fibonacci numbers.} Of course, the choice of Fibonacci numbers were
 based in their generalized occurence in nature; although, we've also 
 tried other sets of numbers, but with no success.
 
 Consider the three Fibonacci numbers below

\begin{equation}
\label{yamashima}
\left\{ F(n) , F(n+1),F(n+k)\right\}.
\end{equation}
where $ n $ is a sufficiently large number and $ k=0,1,2,3,..,N$ ( $ N $ is finite number).

Therefore, we can write the main assumption as 

\begin{equation}
\label{eq8a}
\frac{x(n)}{y(n)+z(n)+w(n)} \propto \frac{F(n)}{F(n+k)},
\end{equation}

\begin{equation}
\label{eq9a}
\frac{y(n)}{x(n)+z(n)+w(n)} \propto \frac{F(n+1)}{F(n+k)},
\end{equation}

\begin{equation}
\label{eq10a}
\frac{z(n)}{x(n)+y(n)+w(n)} \propto \frac{F(n)}{F(n+k)},
\end{equation}

\begin{equation}
\label{eq11a}
\frac{w(n)}{x(n)+y(n)+z(n)} \propto \frac{F(n+1)}{F(n+k)},
\end{equation}

where $ x(n), y(n) , z(n) $ and $ w(n) $ represent the nucleotide frequencies, without
any {\em a priori} association, when the
number of nucleotide bases is $n$, i. e., $ x(n)= \frac{x_n}{n} $, where $ x_n $ stands for
the number of nucleotide $ x $.

It is not straightforward to recognize Chargaff's second parity rule in equations (\ref{eq8a}) -
(\ref{eq11a}). One way to grasp the idea behind the formulas is to note that equations (\ref{eq8a}) and (\ref{eq10a}) are proportional to the same quotient $ \left( \frac{F(n)}{F(n+k)} \right) $,
and the same can be said about equations (\ref{eq9a}) and (\ref{eq11a}). 
In next section, we will show how to get Chargaff's second parity rule from the
above equations.

\subsubsection{Limit Values}

Now, lets impose our second assumption, i. e., that the nucleotide frequencies
tend to limit values when $ n $ is sufficiently large. Mathematically, it can be
written as

\begin{equation}
x=\lim_{n\rightarrow\infty}\frac{x_n}{n}
\end{equation}

\begin{equation}
y=\lim_{n\rightarrow\infty}\frac{y_n}{n}
\end{equation}

\begin{equation}
z=\lim_{n\rightarrow\infty}\frac{z_n}{n}
\end{equation}

\begin{equation}
w=\lim_{n\rightarrow\infty}\frac{w_n}{n}
\end{equation}

It is also necessary to understand what happens with the
quotients $ \frac{F(n)}{F(n+k)} $ and $ \frac{F(n+1)}{F(n+k)} $
when $n\rightarrow\infty $.

Using equation (\ref{eq1}) recursively, it is easy to get the following
recurrence formula

 \begin{equation}
 \label{eq2}
 F(n+k)= F(k) F(n+1)+ F(k-1) F(n). 
 \end{equation}

We are particularly interested in the cases where $ n $, the numbers of bases, is
large, and the quotient of the Fibonacci numbers tends to a limit. 

Mathematically, this can be obtained as follows. Dividing 
(\ref{eq2}) by $ F(n+k) $, we get

\begin{equation}
 \label{eq3}
 1=F(k) \frac{F(n+1)}{F(n+k)}+ F(k-1) \frac{F(n)}{F(n+k)}. 
 \end{equation}

Taking the limit as $n\rightarrow\infty $,

\begin{equation}
 \label{eq4}
 1= F(k) \lim_{n\rightarrow\infty}\frac{F(n+1)}{F(n+k)}+ F(k-1) \lim_{n\rightarrow\infty}\frac{F(n)}{F(n+k)},
 \end{equation}

We define 

\begin{equation}
 \label{eq5}
 \lambda_{1,k}=\lim_{n\rightarrow\infty}\frac{F(n+1)}{F(n+k)}
 \end{equation}

and 

\begin{equation}
 \label{eq6}
 \lambda_{2,k}=\lim_{n\rightarrow\infty}\frac{F(n)}{F(n+k)}
 \end{equation}

Notice that $ \lambda_{1,k} $ and $ \lambda_{2,k} $ are linked to the Golden Ratio by

\begin{equation}
 \lambda_{1,k} = \phi^{1-k}
\end{equation}

and

\begin{equation}
\lambda_{2,k} = \phi^{-k}, 
\end{equation}
respectively.

Thus, the equation (\ref{eq4}) can be written as

\begin{equation}
 \label{eq7}
 1=F(k) \lambda_{1,k}+ F(k-1) \lambda_{2,k}
 \end{equation}
 
Finally, our model can be rewritten as

\begin{equation}
\label{eq8}
\frac{x}{y+z+w}= \lambda_{1,k},
\end{equation}

\begin{equation}
\label{eq9}
\frac{y}{x+z+w}= \lambda_{2,k},
\end{equation}

\begin{equation}
\label{eq10}
\frac{z}{x+y+w}= \lambda_{1,k},
\end{equation}

\begin{equation}
\label{eq11}
\frac{w}{x+y+z}= \lambda_{2,k},
\end{equation}

As noted before, $x$, $y$, $z$ and $w$ are frequencies, so we have 

\begin{equation}
\label{eq12}
x+y+z+w=1.
\end{equation}

Using equation (\ref{eq12}), the equations (\ref{eq8})-(\ref{eq11}), we get

\begin{equation}
\label{eq13}
\frac{x}{1-x}= \lambda_{1,k},
\end{equation}

\begin{equation}
\label{eq14}
\frac{y}{1-y}= \lambda_{2,k},
\end{equation}

\begin{equation}
\label{eq15}
\frac{z}{1-z}= \lambda_{1,k},
\end{equation}

\begin{equation}
\label{eq16}
\frac{w}{1-w}= \lambda_{2,k},
\end{equation}

Equations (\ref{eq13}) and (\ref{eq15}) imply that 

\begin{equation}
\label{eq16.1}
x=z
\end{equation}
and, analogously, equations (\ref{eq14}) and (\ref{eq16}) imply that

\begin{equation}
\label{eq16.2}
y=w
\end{equation}

Equations (\ref{eq16.1}) and (\ref{eq16.2}) are
the Chargaff's second parity rule. 
 
Moreover, an immediate consequence of equations (\ref{eq16.1}), (\ref{eq16.2}) and(\ref{eq12}) is

\begin{equation}
\label{eq17}
 x+y = \frac{1}{2} 
\end{equation}

\subsection{Optimization Problem}

Now, we have three equations in two variables

\begin{eqnarray}
\label{eq17.1a}
\frac{x}{1-x} & = & \lambda_{1,k}  \\
\frac{y}{1-y} & =  &\lambda_{2,k} \\
x+y           & =  &\frac{1}{2} 
\end{eqnarray}
which can be rewritten as

\begin{eqnarray}
\label{eq17.1b}
 x & = &\frac{\lambda_{1,k}}{1+\lambda_{1,k}} \\
 y & = &\frac{\lambda_{2,k}}{1+\lambda_{2,k}}\\
x+y &= & \frac{1}{2} 
\end{eqnarray}
This is a linear system, and, using equations (\ref{eq7}) and 
(\ref{eq17}), it is not difficult to show that it is inconsistent, independently,
of $ k $. In fact, only when $ k\rightarrow\infty $, the system is consistent, but
we are dealing with the cases where $ k $ is finite.

The equation (\ref{eq17}) must be satisfied because $ x $ and $ y $ are frequencies
and, by definition, the equation (\ref{eq12}) must hold. Therefore,
we should try to minimize the difference between $ x $ and $ \frac{\lambda_{1,k}}{1+\lambda_{1,k}} $, and the difference between
$ y $ and $ \frac{\lambda_{2,k}}{1+\lambda_{2,k}} $ under the condition that $ x+y=\frac{1}{2} $.

This is a classical optimization problem, and can be mathematically stated as

\begin{equation}
\label{eq18}
 \min_{x+y=\frac{1}{2}} f_k(x,y), 
\end{equation}

where

\begin{equation}
\label{eq19}
  f_k(x,y)= \left(x- \frac{\lambda_{1,k}}{1+\lambda_{1,k}}\right)^2 + \left(y- \frac{\lambda_{2,k}}{1+\lambda_{2,k}}\right)^2
\end{equation}

This minimization problem is sufficiently easy to solve, because its
objective function  is quadratic and the Jacobian of the constraint
is full rank, therefore the solution exists and is unique \cite{nocedal}. 

In Table 1 we list the solutions to the first $ 8 $ values of $ k $.
It is not difficult to show that $(x,y)\rightarrow (0.25,0.25)$ as
$ k\rightarrow\infty $.

\begin{table}[ht]
 \begin{center}
 %{\tiny
\begin{tabular}{c|l|c|l|c}
      \hline
      k &  $ x $  & $ x\cong$ & $ y $ & $ y\cong$  \\
      \hline
			0 & $\frac{3+\sqrt{5}}{8+4\sqrt{5}} $&0.3090& $\frac{1+\sqrt{5}}{8+4\sqrt{5}}$& 0.1909 \\
			1 & $\frac{3+\sqrt{5}}{8+4\sqrt{5}} $&0.3090& $\frac{1+\sqrt{5}}{8+4\sqrt{5}}$& 0.1909 \\
			2 & $\frac{127+57\sqrt{5}}{420+188\sqrt{5}}$ &0.3027& $\frac{83+37\sqrt{5}}{420+188\sqrt{5}}$ & 0.1972  \\
			3 & $\frac{161+72\sqrt{5}}{550+246\sqrt{5}}$ &0.2927& $\frac{114+51\sqrt{5}}{550+246\sqrt{5}}$& 0.2072 \\
			4 & $\frac{881+392\sqrt{5}}{3126+1398\sqrt{5}}$ &0.2818& $\frac{682+305\sqrt{5}}{3126+1398\sqrt{5}}$& 0.2181 \\
			5 & $\frac{20583+9205\sqrt{5}}{75588+33804\sqrt{5}}$ &0.2723& $\frac{17211+7697\sqrt{5}}{75588+33804\sqrt{5}}$& 0.2276 \\
			6 & $\frac{15908+7070\sqrt{5}}{59665+26683\sqrt{5}}$ &0.2649& $\frac{3(9349+4181\sqrt{5})}{119330+53366\sqrt{5}}$& 0.2350 \\
			7 & $\frac{100793+45076\sqrt{5}}{388045+173539\sqrt{5}}$ &0.2597& $\frac{186459+83387\sqrt{5}}{776090+347078\sqrt{5}}$& 0.2402 \\

\end{tabular}
%}
\end{center}

\begin{center}{ {\tiny Table 1:  Solutions of the optimization problem for different values of $ k $}}
\end{center}
\end{table}

The values of Table 1 are in agreement with the observed frequencies in 
human chromosomes. In the next section, we will present the 
data that supports this mathematical model.

\section{Results}

We've performed a simple experiment using the nucleotide frequencies
in human genome. The human genome data were obtained at NCBI\footnote{National Center for Biotechnology Information. Site (http://www.ncbi.nlm.nih.gov)}. From time
to time new human genome releases are deposited. We've used Build 35.1. It is 
important to note that only partial data is available for each chromosome, i.e.,
there are still missing sections (for example, chromosome 1 is supposed to have
about 263 million bases, but only about 220 million bases were available). 
This information is relevant because it can explain some minor deviations 
from the predicted values.

 \subsection{Human Nucleotide Frequencies}

 This experiment consisted in calculating the nucleotide frequencies
 in all 24  human chromosomes. The results are summarized in the following table.
 
\begin{center}
\tiny
\begin{tabular}[h]{l|c|c|c|c|c}
      \hline
     Chromosome & $ P_A $ & $ P_C $ & $ P_G $ & $ P_T $ & $k$ \\
      \hline
      Chrom 1  & 0.2916 & 0.2080 & 0.2080 & 0.2922 & 3\\
      Chrom 2  & 0.3000 & 0.2003 & 0.2005 & 0.2997 & 2\\
      Chrom 3  & 0.3019 & 0.1980 & 0.1980 & 0.3020 & 2\\
      Chrom 4  & 0.3093 & 0.1905 & 0.1906 & 0.3094 & 1\\
      Chrom 5  & 0.3020 & 0.1974 & 0.1975 & 0.3011 & 2\\
      Chrom 6  & 0.3024 & 0.1975 & 0.1976 & 0.3023 & 2\\
      Chrom 7  & 0.2950 & 0.2040 & 0.2040 & 0.2951 & 3\\
      Chrom 8  & 0.3002 & 0.2001 & 0.2000 & 0.2999 & 2\\
      Chrom 9  & 0.2933 & 0.2067 & 0.2067 & 0.2931 & 3\\
      Chrom 10 & 0.2922 & 0.2074 & 0.2074 & 0.2928 & 3\\
      Chrom 11 & 0.2925 & 0.2072 & 0.2075 & 0.2926 & 3\\
      Chrom 12 & 0.2950 & 0.2040 & 0.2033 & 0.2956 & 3\\
      Chrom 13 & 0.3072 & 0.1922 & 0.1922 & 0.3080 & 1\\
      Chrom 14 & 0.2951 & 0.2034 & 0.2039 & 0.2974 & 3\\
      Chrom 15 & 0.2903 & 0.2101 & 0.2099 & 0.2895 & 3\\
      Chrom 16 & 0.2750 & 0.2040 & 0.2040 & 0.2750 & 4\\
      Chrom 17 & 0.2740 & 0.2261 & 0.2258 & 0.2713 & 5\\
      Chrom 18 & 0.3014 & 0.1982 & 0.1985 & 0.3017 & 2\\
      Chrom 19 & 0.2588 & 0.2403 & 0.2409 & 0.2598 & 7\\
      Chrom 20 & 0.2785 & 0.2194 & 0.2202 & 0.2817 & 5\\
      Chrom 21 & 0.2940 & 0.2040 & 0.2039 & 0.2952 & 3\\
      Chrom 22 & 0.2605 & 0.2398 & 0.2397 & 0.2598 & 6\\
      Chrom X  & 0.3027 & 0.1968 & 0.1967 & 0.3033 & 2\\
      Chrom Y  & 0.3098 & 0.1893 & 0.1889 & 0.3118 & 1\\
      
 \label{tab:T2}
\end{tabular}
\begin{center}{ {\tiny Table 2: Nucleotide Frequencies for all human chromosomes}}
\end{center}
\end{center}

\begin{figure}[ht]
	\begin{center}
	\scalebox{0.7}
	{
		\includegraphics{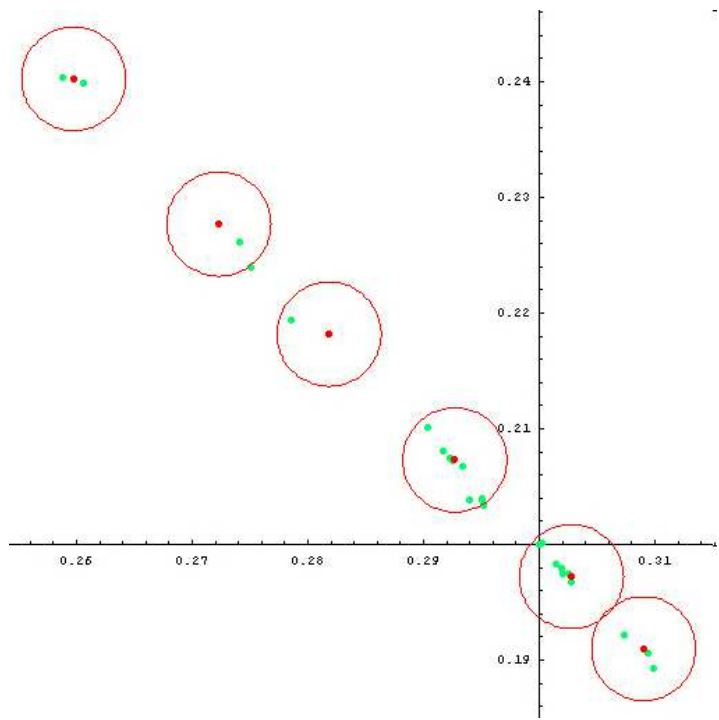}
	}	
	\end{center}
	\caption{In red dots, the solutions of the optimization problem for different values of
	 $ k $. In green, the points $ (P_A , P_C) $ for each one of the human chromosomes.}
	\label{fig:fibsol}
\end{figure}

 The nucleotide frequencies are clustered around  
 the predicted values of Table 1. In Figure \ref{fig:fibsol},
 we have in red the solutions of the optimization problem for different
 values of $ k $, and in green the points $ (P_A , P_C) $ for each one
 of the human chromosomes. The red circles have their centers 
 in the solutions of the optimization problem and they have the 
 same radius equal to $ 0.005$. 
 
 It is interesting to note that all the
points $ (P_A , P_C) $ are near ( less than $ 0.005$)
to the predicted values.

 The average values are $ \mu_{P_A} = 0.292 $, $ \mu_{P_C}=0.207 $,
 $ \mu_{P_G}=0.207 $ and $ \mu_{P_T} = 0.292 $, which are close
 to the optimization's solution when $ k=3 $.

\section{Conclusion}

Using Chargaff's second parity rule and
assuming that the nucleotide frequencies tend
to limit values when the number of nucleotide bases
is sufficiently large, we've described a mathematical model that
predicts the limit values of the human nucleotide
frequencies with great accuracy. It is also interesting to
note that the limit values are the results of an optimization
problem, and it is commonly found in many phenomena in nature.

If our two hypotheses hold and our mathematical model 
is correct, then it is possible to make the following conjecture: the 
noncoding DNA regions play a major rule in the ``optimization
process'' to reach the limit values predicted in our mathematical
model. This conjecture is based on the fact that  about
$ 97 \% $ of human genome is believed to be noncoding.

\subsection*{Acknowledgements}

We are grateful to Dr. Bernard Maigret from Henry Poincar\'e University (Nancy - France), Dr. Robert Giegerich from Bielefeld University ( Bielefeld - Germany) and Dr. Nir Cohen from Campinas
State University (Campinas - Brazil) for their valuable comments on our manuscript.

\end{document}